\documentclass[preprint, review, 12pt]{elsarticle}
\usepackage[utf8]{inputenc}
\usepackage{amsmath}
\usepackage{nicefrac}
\usepackage{setspace}
\usepackage{booktabs}
\usepackage{icomma}
\usepackage{comment}
\usepackage[T1]{fontenc}
\usepackage{lmodern} % monospace font
\usepackage[scale=0.89]{tgheros} % Helvetica is too big
\usepackage{Baskervaldx} % tosf in text, tlf in math;; use [osf] to get old time numbers -- better without old time numbers for %tech document
\usepackage[baskervaldx,cmintegrals,bigdelims,vvarbb]{newtxmath} % math italic letters from Baskervaldx
\usepackage[cal=boondoxo]{mathalfa} % mathcal from STIX, unslanted a bit

\newcommand{\mybib}{\vspace{0.05 in}\setlength{\hangindent}{0.35in}\noindent\samepage }

\journal{arXiv}
\author{Rajesh P. Narayanan\fnref{inst1}}
\affiliation[inst1]{organization={Department of Finance, Louisiana State University},%Department and Organization
           city={Baton Rouge},
           postcode={70803}, 
           state={LA},
           country={USA, rnarayan@lsu.edu}}

\author{R. Kelley Pace\fnref{inst2}}
 \affiliation[inst2]{organization={Department of Finance, Louisiana State University},
           country={U.S.A}}

\title{\textbf{Can the Nexus of Scaling Laws Coupled with  Constant or Variable Elasticity of Substitution Predict AI and Other Technology Adoption? }}
\begin{document}      

\begin{frontmatter}

\begin{abstract}

Emergent technologies such as solar power, electric vehicles, and artificial intelligence (AI) often exhibit exponential or power function price declines and various ``S-curves'' of adoption. We show that under CES and VES utility, such price and adoption curves are functionally linked. When price declines follow Moore's, Wright's and AI scaling ``Laws,'' the S-curve of adoption is Logistic or Log-Logistic whose slope depends on the interaction between an experience parameter and the elasticity of substitution between the incumbent and emergent good.  These functional relations can serve as a building block for more complex models and guide empirical specifications of technology adoption.  
\vspace{0.1in}
\end{abstract}

\begin{keyword}

Wright's Law \sep Moore's Law \sep AI scaling laws \sep S-curve adoption \sep CES \sep VES
\sep climate change \sep elasticity of substitution.
%% \renewcommand{\PACS}{\text{JEL classification}\enspace}

\renewcommand{\PACS}{\text{JEL classification}}
\PACS: {O30 \sep O31 \sep O33  \sep Q55 \sep D24}
\end{keyword}

\end{frontmatter}

\begin{comment}

\clearpage

\section*{Significance}
\large
The adoption of emerging technologies like solar power, electric vehicles, and artificial intelligence follows certain predictable patterns. This research helps explain why these technologies often experience exponential price declines and then surge in adoption, forming ``S-curve'' growth patterns. By using an economic concept called CES utility, we show how well-known price reduction trends, such as Moore's and Wright's Laws, are directly linked to these adoption patterns. Understanding this connection helps policymakers, companies, and investors better anticipate when new technologies will become widely used and which factors will influence their speed of adoption.

\end{comment}

%\maketitle
\setstretch{1.25}

\large
\section{Introduction}

%Emergent technologies such as solar panels, batteries, electric vehicles, and artificial intelligence (AI) may  affect the future of the climate and the economy. Emergent technology producers must forecast  the speed of adoption of these  over a substantial time horizon to justify near-term capital expenditures.

Emergent technologies such as solar power, batteries, electric vehicles, and artificial intelligence (AI) display two empirical regularities.  First, they exhibit exponential or power function price declines or quality improvements over time. Second, the adoption of these technologies typically follows an S-curve pattern. This raises the question: what governs the relationship between the price decline curve and the adoption curve?

We show using a constant elasticity of substitution (CES) and a variable elasticity of substitution (VES) utility function (Arrow et al., 1961; Dixit et al., 1977),  if  price declines follow Wright's Law, then the adoption follows a log-logistic curve; if they follow Moore's Law, the adoption follows a logistic curve.  The slope of these adoption S-curves depends on an experience or learning  parameter, and the elasticity of substitution between the incumbent and emergent good (or service).
%These functional relations can provide a building block for other models and suggest empirical specifications of technology adoption.

The experience or learning parameter in the price functions  governs the potential rate prices could fall as a function of cumulative production (Wright's Law) or over time (Moore's Law). These reductions come from experience or from learning with more production or time spent with the technology. The parameters can vary across technologies. For example, although solar prices have fallen over time per panel watt/hour, the price of secondary storage per gigabyte has declined even faster.

%The CES utility function, which includes the Cobb-Douglas utility function,   provides the most common way to model consumer choice. 

%actually, the elasticity of substitution is more fundamental than CES

The elasticity of substitution parameter represents the  sensitivity of consumer choice to the relative prices of products in the CES utility function (the most common specification in consumer choice) as well as the VES utility function. If consumers (who might be individuals, businesses, governments, or other organizations) are extremely sensitive to relative prices (high elasticity of substitution), price declines from technological progress might quickly lead to adoption of emergent goods or services. Alternatively, if they are insensitive to relative prices (low elasticity of substitution), price declines from technological progress may lead to  slow growth in the emergent good or service. Differences in the elasticity of substitution could arise in several ways. First, frictions (i.e., existing contracts, established relationships, information costs) can lead to a low elasticity of substitution in the short run, but this can rise over time. Second, differences between the incumbent and emergent good may reduce the elasticity of substitution (e.g., electric and gas cars are transportation, but drive differently). Third, standards, regulation, and infrastructure could further affect the elasticity (e.g., concrete substitutes may require additional testing and certification). 

%Note, the CES utility function, which includes the Cobb-Douglas utility function,   provides the most common way to  . 

Currently, many works support empirical models of price declines such as Moore's or Wright's Laws (Arrow, 1962; Nagy et al., 2013; Magee et al., 2016; Lafond et al., 2018; Triulzi et al., 2020; Singh et al., 2021; Way et al., 2022). Moore's Law refers to an exponential decrease in price over time (log-linear in time) while Wright's Law refers to price declining as a power function of total production over time (log-linear in total production-to-date). Both of these ``scaling laws'' seem to empirically outperform the alternative economic construct of economies-of-scale. 

Similarly, AI scaling laws (Kaplan et al., 2020; Hoffmann et al. 2022) show accuracy improvements (reductions in quality-adjusted price) as a power function of computation.\footnote{As an indication of the influence of these works as of November 2024 both of these papers have over 2,000 Google Scholar citations with over 1,000 each in 2024 alone.} ``The consistent and predictable improvements from scaling have led AI labs to aggressively expand the scale of training, with training compute expanding at a rate of approximately 4x per year'' (Sevilla et al., 2024). As an indication of the importance of the scaling laws motivating investment in AI, the CEO of Google DeepMind, Demis Hassabis (winner of the 2024 Nobel Prize in Chemistry) indicated that Google would be spending \$100 Billion or more on AI (Seal, 2024).

%Can AI Scaling Continue Through 2030?

Also, many works have explored the adoption of emerging goods relative to incumbent goods.\footnote{Griliches (1957) examined uptake of hybrid versus open-pollinated seeds, Gr\"{u}bler et al., (1999) discuss cars versus horses, Way et al. (2022) forecast batteries as well as wind, and  Li et al. (2023) study residential solar adoption.} Empirical studies show an ``S-curve'' of adoption where the emerging good begins with a higher price and  lower market share, but through price declines or quality improvements gains product  share from an incumbent good based on a  mature technology. This S-curve has often  been modeled with a logistic function (Griliches, 1957; Way et al., 2022; Creutzig et al., 2023). 
   
The  scaling law curves and the adoption S-curves have been studied separately, although the connection between these two  has been acknowledged and used in either informal or complicated ways (Gr\"{u}bler et al., 1999; Odenweller,  2022;  Nijsse et al., 2023). Many of these works come from areas other than economics (Bass et al., 1994; Haegel and  Kurtz, 2023; Li et al., 2023).  Using economic constructs such as utility and elasticity of substitution to link price trajectories and  S-curves may offer additional insights and may help predict future prices and adoption that can inform investment decisions, the design of innovation policy in areas such as AI, renewable energy or other emerging fields (Agrawal et al., 2019) and the trade-off between human and AI labor. 

In this paper, we examine the link between these empirical regularities in a  constant elasticity of substitution (CES) utility framework. Specifically, we show that choices made based on CES utility with a budget constraint given Moore's, Wright's, and AI scaling laws lead to logistic (Moore's Law) or log-logistic (Wright's Law, AI scaling laws) S-curves. We show that these relations can also hold in a VES framework.

We provide an example involving AI scaling laws to demonstrate the usefulness of these functional relations in understanding AI-human labor trade-offs. More generally, these findings  have implications for transitions between incumbent and emergent goods and services as well as the associated  customers,  regulation, and infrastructure. In particular, the reduction of employees at incumbent firms and the gain of employees at emergent firms may lead to issues with the transition.

\section{CES Utility and Shares}\label{ces_section}

We begin with a CES utility function  for two goods $x$ and $y$ in \eqref{cesu}  where $\alpha$ specifies the relative preference for good $x$, $(1-\alpha)$ specifies the relative preference of good $y$, $M$ is income, and $\rho$ is a function of the elasticity of substitution $\sigma$ in \eqref{constraints}.   Special cases for $\rho$ include Cobb-Douglas ($\rho=0$ which implies $\sigma=1$), perfect substitutes ($\rho \rightarrow 1 \text{ or } \sigma \rightarrow \infty$), and perfect complements ($\rho \rightarrow -\infty \text{ or } \sigma \rightarrow 0$).

\begin{align}
U(x, \ y)&=\left [ \alpha \cdot  x^{\rho} +(1-\alpha) \cdot y^{\rho} \right ]^{1/\rho} \!,  \, \alpha \in (0, 1), \, \rho \in (-\infty, 1) \label{cesu}\\
& \text{s.t.} \ \ p_{x}\cdot x+p_{y} \cdot y=M, \ \ \sigma=(1-\rho)^{-1}\label{constraints}
\end{align}

\noindent
The demand equations associated with the CES utility function above (Silberberg and Suen, 2017, p. 359--360) appear in \eqref{demandx}--\eqref{cesconstant}.

\begin{align}
x&=D \cdot \alpha^{\sigma}\cdot p_{x}^{-\sigma}\label{demandx}\\
y&=D \cdot (1-\alpha)^{\sigma} \cdot p_{y}^{-\sigma}\label{demandy}\\
D&=(\alpha^\sigma \cdot p_{x}^{1-\sigma}+(1-\alpha)^{\sigma} \cdot p_{y}^{1-\sigma})^{-1} \cdot M \label{cesconstant}
\end{align}

\noindent
Our focus is on the relative fraction $S_{x}$  in the market as shown in the product share equation for good $x$ in \eqref{shareces}.

\begin{align}\label{shareces}
S_{x}&=\frac{x}{x+y}=\frac{1}{1+y/x}=\frac{1}{1+\left [\frac{1-\alpha}{\alpha}\right]^{\sigma} \cdot \left (\frac{p_{x}}{p_{y}}\right)^{\sigma}}
\end{align}

\noindent
Because the incumbent good has a stable technology and because the focus is upon relative prices, we set the price of the incumbent good $p_{y}$ to $1$ which leads to \eqref{share}.

\begin{align}
S_{x}&=\frac{1}{1+\left [\frac{1-\alpha}{\alpha}\right]^{\sigma} \cdot p_{x}^{\sigma}}\label{share}
\end{align}

\section{Moore's Law and Wright's Law Imply Logistic and Log-Logistic S-Curves}

In examining  price curves over time for a number of different products, the literature (Arrow, 1962; Nagy et al., 2013; Magee et al., 2016; Lafond et al., 2018; Triulzi et al., 2020; Singh et al., 2021; Way et al., 2022) often finds greater empirical support for   Moore's Law and Wright's Law functional forms than alternatives such as economies-of-scale which depend only upon current production levels.  Moore's Law in \eqref{Moore} models price $p_{x}(t)$ as falling exponentially over time at rate $m$ while Wright's Law in \eqref{Wrights} models price as falling with cumulative production of good $x$ denoted as $X(t)$ as a function of a parameter $s$. The parameter $B$ is the first unit's  price.  

\begin{align}
p_{x}(t)&=B \cdot \exp(-m \cdot t)\label{Moore}\\
p_{x}(t)&=B \cdot X(t)^{-s} \label{Wrights}\\
& B, s, m > 0, \ t \ge 0 \label{cons2}
\end{align}

\noindent

Specifically, we examine $S_{x}(t)_{M}$ for Moore's Law in \eqref{Fmoores} and $S_{x}(t)_{W}$ for Wright's Law  in \eqref{Fwrights}.

\begin{align}
S_{x}(t)_{M}&=\frac{1}{1+  K^{\sigma} \cdot \exp(-\sigma \cdot m \cdot t)}\label{Fmoores}\\
S_{x}(t)_{W}&=\frac{1}{1+K^{\sigma}  \cdot X(t)^{-\sigma  \cdot s}}\label{Fwrights}\\
K&=\left  [\frac{1-\alpha}{\alpha}\right] \cdot B \label{bigcon}
\end{align}

By construction $S_{x}(t)$ lies in $[0, \ 1]$ and even though the model does not involve random variables, this suggests that $S_{x}(t)$ might have the forms associated with a cumulative density function (cdf). In fact, Moore's Law suggests a logistic cdf model as shown in \eqref{logistic} with mean parameter $\mu$ and scale parameter $\lambda$ in \eqref{logisticscale}.

\begin{align}
S_{x}(t)_{M}&=\frac{1}{1+   \exp(-\sigma \cdot m \cdot t +\sigma \cdot \ln(K))}\label{Fmoores2}\\
F(t)_{M}&=\dfrac{1}{1+\exp(-(t-\mu)/\lambda)}\label{logistic}\\
\mu&=\dfrac{\ln(K)}{ m} , \ \lambda= \dfrac{1}{\sigma \cdot m} \label{logisticscale}
\end{align}

The logistic form for Moore's Law suggests rewriting the share of the emergent good using Wright's Law as an exponential Moore's Law in \eqref{expwright} where the logarithm of cumulative production $\ln(X(t))$ follows a logistic distribution.

\begin{align}
S_{x}(t)_{W}&=\frac{1}{1+ \exp( -\sigma \cdot s \cdot \ln(X(t))+\sigma \cdot \ln(K))}\label{expwright}
\end{align}

\noindent
Therefore,  the log-logistic   provides the relevant distribution for Wright's Law (aka,  Fisk  distribution in economics).\footnote{Fisk (1961, eq. 7a) with division of numerator and denominator by $(t/t_{0})^{\alpha}$.} Beginning with the Wright's Law share $S_{x}(t)_{W}$ in \eqref{Fwrights2b} this leads to the log-logistic cdf for $X(t)$ in \eqref{Wrights3} where the definitions for $\beta$ and $\lambda$ appear in \eqref{alphadef}. Note, the distinction between Moore's Law and Wright's Law collapses when $X(t)$ grows exponentially (Sahal, 1979). 

%It is also a Pareto(III) distribution (Arnold et al., 1986, p. 404).

%Note, that these are related since the Pareto (III) distribution with a mean parameter of 0 is sometime termed a log-logistic distribution (Barry, 2014, eq. 14).

\begin{align}
S_{x}(t)_{W}&=\frac{1}{1+[K^{-(1/s)}  \cdot X(t)]^{-\sigma  \cdot s}}\label{Fwrights2b}\\
F(t)_{W}&=\dfrac{1}{1+(X(t)/\lambda)^{-\beta}}\label{Wrights3}\\
\beta&=\sigma \cdot s, \ \lambda=K^{(1/s)}\label{alphadef}
\end{align}

%Alternative  S-curves  emerge  when employing other forms of cost declines modeled using reliability or survivor functions (defined as $1-F(t)$) such as those from the Gompertz, Weibull, and Pareto distributions. 

\section{An AI Scaling Law Example}

Suppose a purely human created product exists at $p_{H}$ with an associated level of quality that does not vary over time.  For convenience,  let the quality-adjusted price $p_{H}$ equal 1. We assume that the pecuniary price of the AI-created product equals a constant $B$ and at $t=0$ has lower quality than the human product. 

However, AI-created products may show quality-adjusted price declines over time. Based on the Kaplan et al. (2020) and Hoffmann et al. (2022) scaling laws, the  loss or inaccuracy from using AI, as measured in this case by the negative log-likelihood per character, $-L(C(t))/n$,   declines with the amount of compute $C$ available via a power law as shown in \eqref{Delta} which involves a proportional constant  $\kappa$ and an exponent $\delta$ governing the rate of decline. The term $-L(C(\infty))/n$ represents the irreducible minimum loss per character that even an infinite amount of computation would not lower.\footnote{Recently scaling has extended to ``test time'' compute in addition to the pretraining compute. Chollet (2024) shows a graph that indicates that the performance on the Open AI o3 model seems to increase log-linearly with compute (power law) and this continues for approximately three orders of magnitude relative to the current best public model, o1. Thus,  scaling ``laws'' seem to hold when going to test time compute.}  We simplify \eqref{Delta} slightly in \eqref{Delta2}.

\begin{align}\label{Delta}
-L(C(t))/n&\approx -L(C(\infty))/n + \kappa \cdot   C(t)^{-\delta}, \ C(t),  \delta,  \kappa,  t >0\\
\Delta_{\text{AI}}(C(t))&\approx \kappa \cdot   C(t)^{-\delta} \label{Delta2}
\end{align}

\noindent
As the  excess AI Loss $\Delta_{\text{AI}}(C(t))$ in \eqref{Delta} declines (accuracy increases), the quality-adjusted price $p_{\text{AI}}(t)$ should also decline. 

However, the quality-adjusted price may not decline at the same rate as the excess AI Loss $\Delta_{\text{AI}}(C(t))$. Using AI benchmarks as a surrogate for individual AI products, shows that as excess loss declines with increases in computation, benchmark performance may rise at different rates (Ott et al., 2022, Figure~3, p. 3; Maslej et al., 2024, Chapter 2, p. 9) relative to the rate of excess loss  $\Delta_{\text{AI}}(C(t))$ over time.

For additional context, Meta (2024) recently introduced Llama 3.1 with 405 billion parameters (Llama 3.1 405B) and discuss scaling laws as part of model development. Their Figure~4 (Meta, 2024, p. 9) shows both the calculated loss  as well as the performance on the ARC Challenge benchmark (AI2 ARC) as a function of computation. We transferred the points on the graphs into data (16 observations) and ran a linear regression on the loss (as measured by negative log-likelihood per character) as well as a logit regression on the benchmark (relative accuracy from 0 to 1). The single non-constant regressor for both regressions was the $\log_{10}(C)$. The estimated coefficient for the linear regression on loss was $-0.049$ or $\hat\delta=0.049$. This closely agrees with the result in Kaplan (2020) of 0.05 for $\hat\delta$. The estimated coefficient for the logit regression on the benchmark accuracy was $0.511$. All the coefficients on $\log_{10}(C)$ were highly significant with $t$ statistics of $-35.1$ for OLS and 14.8 for logit with an $R^{2}$ of 0.988 for OLS and a McFadden pseudo-$R^{2}$ of 0.966 for logit.\footnote{Estimates came from Matlab routine `glmfit' with logit link and normal errors. A probit link gave similar marginal effects. We measured marginal effects at 0.01 intervals from 18.778 (minimum) to 25.580 (maximum) of $\log_{10}(C)$ as given by the data which gave 681 points. These 681 marginal effect estimates formed the basis for the marginal effects percentiles and mean.}
%The $R^{2}$ for the linear regression on loss and the pseudo $R^{2}$ for the logit regression on accuracy were both over 0.99. 
With regard to the logit regression, the important quantities are the marginal effects, the derivatives of estimated benchmark accuracy with respect to the explanatory variable $\log_{10}(C)$ for different levels of $\log_{10}(C)$. The  marginal effects at the 10, 25, 50, 75, 90 percentiles were 0.069, 0.086, 0.106, 0.122, and 0.127. The average marginal effect was 0.102. Based on the marginal effect estimates, an order of magnitude increase in $C$ results in loss declining by approximately five percent and the benchmark accuracy increasing by approximately 10 percent.

Given this background, we assume that the quality-adjusted price $p_{\text{AI}}(t)$ in \eqref{WrightC} will also follow a power law of excess loss from \eqref{Delta2} governed by a decline  at some strictly positive rate $\gamma$ with a strictly positive proportionality constant $\kappa$, as we only wish to claim a qualitative relation between AI loss and quality. This power law modification for AI specific task performance was chosen for simplicity and other forms could exist.

\begin{align}\label{WrightC}
p_{\text{AI}}(t)&\approx B \cdot \left( \Delta_{\text{AI}}(C(t)) \right)^{\gamma}= B \cdot \kappa^{\gamma} \cdot C(t)^{-\delta \cdot \gamma}, \ B,  \gamma>0 
\end{align}

\noindent
To give some idea of possible magnitudes, based on the  results above, the regression on the loss estimates $-\delta$ while the marginal effects from the logit regression on the benchmark estimate $-\delta \cdot \gamma$. If we approximate and say that $\delta \sim 0.05$ and that $\delta \cdot \gamma \sim 0.10$, then $\gamma \sim 2$. That is, the general scaling reduces the loss at a rate of 0.05 while the benchmark accuracy, a proxy for quality, improves at twice that rate for a total rate of quality-adjusted price decline of 10\% for each factor of 10 increase in $C$. 

To provide more perspective to these results, based on  Figure~2.1.16 from  Maslej et al. (2024, Chapter 2, p. 9), some benchmarks such as competition level mathematics (MATH) increased rapidly from approximately 8\% at the beginning of 2021 to around 96\% at the end of 2022. The MMLU benchmark went from around 38\% at the beginning of 2019 to around 100\% by the end of 2023. In contrast, ImageNet Top-5 made slower progress as it went from a relatively high 90\% in 2012 to 105\% in 2023. These percentages are relative to a human level performance of 100\%. The ARC-AGI benchmark currently serves as an example of a  benchmark that shows slower progress.  In summary, $\gamma$, the parameter governing task-specific performance, could be less than or greater than 1. Tasks where $\gamma>1$ will be more likely to show greater drops in quality-adjusted prices.

%The ARC-AGI grand prize of one-million dollars will be awarded to a score of 85\% or better (given some other constraints).  The current high level in 2024 of 43\%. The organizers (arcprize.org) believe that current AI systems will only show slow increases in this benchmark over time. 

%Ott et al., 2022, Figure 3, p. 3;

As compute $C(t)$ rises over time $t$, the quality-adjusted price $p_{\text{AI}}(t)$ in \eqref{WrightC} falls. This power law relation between compute $C(t)$ and quality-adjusted price $p_{\text{AI}}(t)$  represents a form of Wright's Law based on computational capacity $C(t)$ rather than  historical production. If $C(t)$ grows at an exponential rate then this reduces to a form of Moore's Law.

%, although current computational capacity $C(t)$ reflects the accumulated experience of computation. 

In an industry with an incumbent producer (humans) selling a product at a quality-adjusted price of 1 ($p_{H}=1$) and an emergent producer (AI) selling a product at a quality-adjusted price of $p_{\text{AI}}(t)$, what will happen to the market share of humans versus AI? Following the earlier development of the share equation \eqref{share} and substituting $p_{\text{AI}}(t)$ for $p_{x}$ yields \eqref{sharea1} and \eqref{consai}  that results in  a log-logistic S-curve of adoption in \eqref{shareai2}.

\begin{align}
S_{\text{AI}}(t)&=\frac{1}{1+\left [\frac{1-\alpha}{\alpha}\right]^{\sigma} \cdot p_{\text{AI}}(t)^{\sigma}}\label{sharea1}\\
K&=\left [\frac{1-\alpha}{\alpha}  \right] \cdot B   \label{consai}\\
S_{\text{AI}}(t)&=\frac{1}{1+ K^{\sigma} \cdot \kappa^{\sigma\cdot\gamma} \cdot C(t)^{-\sigma \cdot \delta \cdot \gamma }} \label{shareai2}
\end{align}

\noindent
This simple example highlights key factors that could increase the share of AI. Artificial intelligence will more easily disrupt highly substitutable  human products or tasks (high $\sigma$) when using models that learn quicker (high $\delta$) for tasks or products where a given decrease in AI loss  results in a larger decrease in the quality-adjusted price (high $\gamma$). Of particular interest are products or tasks where the product of these parameters $\sigma \cdot \delta \cdot \gamma$  have higher (lower) values leading to rapid (slow) adoption.

%<1$ which would lead to AI making slower incursions into human tasks. For example, regulation may prevent AI from directly prescribing medications (low $\sigma$). However, physicians may use AI to increase their productivity and quality (e.g., AI screening for drug interactions).

%and humans as complements.  Tasks with those characteristics as well as potential hybrids of the human and AI product will leave a greater role for the human producer. 
%If $\sigma \cdot \delta \cdot \gamma >1$, AI and humans are substitutes which suggests a diminished human role.

To provide a current context, consider the  progress of language translation by AI. A recent survey by the Society of Authors in 2024 shows that one-third of translators are losing work to AI. Also, Aslanyan (2024) reports that the Swedish publisher Lind \& Co ``only uses machine translation for genres such as crime and romance.''
At the moment, AI is still not thought to have sufficient quality to handle more complicated literary works. If AI compute $C$ continues to rise over time, through expansion of investment (nearly octupled in 2023, Maslej et al., 2024, p. 5) or by secular technological increases in coding efficiency and algorithmic innovations,  other genres may fall to AI and AI translation will follow a log-logistic S-curve of adoption.

\section{A Variable Elasticity Approach}\label{ves}

The constant elasticity of substitution (CES) approach followed above  provides one way of viewing the trade-off between AI and human conducted  or products. This approach combines prices and quality changes into a quality-adjusted price. However, we now examine a variable elasticity of substitution (VES) approach which separately specifies price and allows the quality to affect the elasticity of substitution. Separating these two important features (price, quality) may aid in understanding and interpreting the trade-off between AI and human conducted tasks. Interestingly,  the CES and the VES approaches can both arrive at a log-logistic form for the AI adoption curve. 

To motivate the VES approach,  more computation increases the abilities of AI and therefore allows  AI work to more easily substitute for human work. In other words, the elasticity of substitution varies with computation and to capture this notion we adopt a variable of elasticity (VES) approach. Rather than examine a quality-adjusted price as done previously, we now treat $p_{\text{AI}}$ as a constant. Specifically, we assume $p_{\text{AI}}<p_{H}=1$, $K>1$, and allow the elasticity of substitution $\sigma$ to become a function of the benchmark sensitivity parameter $\gamma$, the learning parameter $\delta$,  and the level of computation $C(t)$ scaled by $C(0)$. We also assume that computation grows monotonically over time so that $C(t) \ge C(0)$.   Specifically,  we assume that the elasticity of substitution rises with the log of computation as modified by the benchmark sensitivity $\gamma$ and the learning parameter $\delta$ as shown in \eqref{spec}.

\begin{align}\label{spec}
\sigma( \gamma, \delta,  \, C(t)) &=   \gamma \cdot \delta \cdot \ln(C(t)/C(0))
\end{align}

\noindent
We assume that the product of $ \gamma \cdot \delta \in (0, \,1)$. If $C(t)=C(0)$, $\sigma( \gamma, \delta,  \, C(t)) =0$, but rises with more computation. Therefore, the share equation takes the form
\begin{align}
S_{AI}(t) &= \frac{1}{1+K \cdot p_{AI}^{\sigma(\gamma, \, \delta, \, C(t)/C(0))}}. \label{share_eq_start}
\end{align}

\noindent
 Substitution of \eqref{spec} into \eqref{share_eq_start} leads to \eqref{sub_sigma}.
 
\begin{align}
S_{AI}(t) &= \frac{1}{1+K \cdot p_{AI}^{ \gamma \cdot \delta \cdot  \ln(C(t)/C(0))}}. \label{sub_sigma}
\end{align}

%\noindent
%Via the exponentiation property,

%\begin{align}
%a^{b \ln(x)} &= x^{\ln(a) b}.
%\end{align}

%\noindent
%Using this with $a = p_{AI}$ and $b = \sigma \cdot \gamma \cdot \delta$ yields \eqref{exp_step}.

\noindent
We now manipulate the term involving $p_{\text{AI}}$ to arrive at \eqref{exp_step}.

\begin{align}
(p_{AI})^{\gamma \cdot \delta \cdot \ln(C(t)/C(0))} &= \exp \left(\ln \left((p_{AI})^{\gamma \cdot \delta \cdot \ln(C(t)/C(0))} \right) \right) \nonumber \\
&= \exp(\gamma \cdot \delta \cdot \ln(C(t)/C(0)) \cdot \ln(p_{AI})) \nonumber \\
 &= \left(\dfrac{C(t)}{C(0)} \right)^{ \gamma \cdot \delta \cdot \ln(p_{AI})}. \label{exp_step}
\end{align}

\noindent
Substitution of \eqref{exp_step} back into the share equation \eqref{sub_sigma} yields \eqref{after_exp},
\begin{align}
S_{AI}(t) &= \frac{1}{1 + K \cdot \left(\dfrac{C(t)}{C(0)} \right)^{ \gamma \cdot \delta \cdot \ln(p_{AI})}}. \label{after_exp}
\end{align}

\noindent
where \eqref{after_exp} represents  a log-logistic form. At level of computation $C(0)$, $S_{\text{AI}}=1/(1+K)$. For $p_{\text{AI}}<1$,  as previously assumed, $\ln(p_{\text{AI}})<0$, and the exponent of $C(t)/C(0)$ is negative since $\gamma \cdot \delta >0$, as previously assumed. The AI share $S_{AI}$ can grow if either $C(t)/C(0)$ becomes large or $p_{\text{AI}}$ becomes small.

\section{VES Extensions}\label{extension}

The share equation \eqref{after_exp} derived in  Section~\ref{ves} conditions on $p_{\text{AI}}$ and $C(t)$. In this section we explore specifying the trajectory of marginal costs of AI which leads to a trajectory of the price of AI and we specify a relation between AI computation and price of AI to arrive at a trajectory for $C(t)$. These assumptions allow for further exploration into AI adoption ($S_{AI}(t)$).

First, we specify the profits from AI on an industry-wide basis in \eqref{industry} where we assume a constant addressable market $A$ and $\theta(t)$ represents the marginal cost of buying an additional unit of $C$.

\begin{align}\label{industry}
\pi_{\text{AI}}(t)&=A \cdot S_{AI}(t) -\theta(t) \cdot C(t)
\end{align}

\noindent
Therefore, $\theta(t)$ represents the marginal cost. For many years the cost of computation has fallen and we model this as an negative exponential type of relation in \eqref{explike} where $\tau$ is a decline parameter and $a$ allows the marginal cost to fall following a super (a>1), exponential ($t=1$), or sub exponential ($t<1$) trajectory.

\begin{align}\label{explike}
\theta(t)&=\exp(-\tau \cdot t^{a})
\end{align}

We assume that in the long-run the price of AI will match the marginal cost as in \eqref{comp}.

\begin{align}\label{comp}
p_{\text{AI}}(t)&=\theta(t)=\exp(-\tau \cdot t^{a})
\end{align}

We now introduce an endogenous relationship between computation $C(t)$ and price $p_{\text{AI}}(t)$. We do not wish to specify whether the computation responds to price or \textit{vice-versa}. Therefore, we model this as an implicit equation in \eqref{compute_price_relation} where $f$ and $g$ are positive constants. Note, the inverse relation between price of AI and computation.

\begin{align}\label{compute_price_relation}
0&=\ln\left(\frac{C(t)}{C(0)}\right) - f + g \cdot \ln(p_{\text{AI}}(t))
\end{align}

\noindent
 Given \eqref{compute_price_relation} and \eqref{comp}, we  express computation $C(t)$ as  \eqref{computation_growth}.

\begin{align}\label{computation_growth}
C(t) &= C(0) \cdot \exp\left(f + g \cdot \tau \cdot t^{a}\right).
\end{align}

\noindent
Substituting \eqref{computation_growth} into \eqref{spec}, the elasticity of substitution appears in \eqref{sigma_updated}.

\begin{align}\label{sigma_updated}
\sigma(\gamma, \delta, C(t)) &= \gamma \cdot \delta \cdot \ln\left(\frac{C(t)}{C(0)}\right) \nonumber \\
&= \gamma \cdot \delta \cdot \left(f + g \cdot \tau \cdot t^{a}\right).
\end{align}

\noindent
Now, substituting \eqref{sigma_updated} into the share equation \eqref{sub_sigma}, we obtain \eqref{after_exp_new}.

\begin{align}
S_{AI}(t) &= \frac{1}{1 + K \cdot p_{\text{AI}}^{\gamma \cdot \delta \cdot (f + g \cdot \tau \cdot t^{a})}} \nonumber \\
&= \frac{1}{1 + K \cdot \left(\exp(-\tau\cdot t^{a})\right)^{\gamma \cdot \delta \cdot (f + g \cdot \tau \cdot t^{a})}} \nonumber \\
&= \frac{1}{1 + K \cdot \exp\left(-\gamma \cdot \delta \cdot \tau \cdot (f\cdot t^{a} + g \cdot\tau \cdot t^{2a})\right)}. \label{after_exp_new}
\end{align}

\noindent
This can be further simplified to \eqref{after_exp_final}.

\begin{align}
S_{AI}(t) &= \frac{1}{1 + K \cdot \exp\left(-\gamma \cdot \delta \cdot \tau \cdot f \cdot t^{a} - \gamma \cdot \delta \cdot \tau^{2} \cdot g \cdot t^{2a}\right)}. \label{after_exp_final}
\end{align}

\noindent
Taking the log-odds ratio (logit) of \( S_{AI}(t) \), we have \eqref{logit_updated}.

\begin{align}\label{logit_updated}
\ln\left(\frac{1 - S_{AI}(t)}{S_{AI}(t)}\right) &= \ln(K) - \gamma \cdot \delta \cdot \tau \cdot f \cdot t^{a} - \gamma \cdot \delta \cdot \tau^{2} \cdot g \cdot t^{2a} 
\end{align}

\noindent
We reorganize \eqref{logit_updated} in \eqref{logit_updated3} to improve interpretability.

\begin{align}
\ln\left(\frac{1 - S_{AI}(t)}{S_{AI}(t)}\right) &= \ln(K) -   (\tau \cdot t^{a}) \cdot (\gamma \cdot \delta)\cdot (f + g \cdot \tau \cdot t^{a}) \label{logit_updated3}\\
&=\ln(K) -  (\tau \cdot t^{a}) \cdot \sigma(\gamma, \delta, C(t)) \label{logit_updated4}
\end{align}

In \eqref{logit_updated3} the term  ($\tau \cdot t^{a}$) governs the rate at which AI becomes cheaper over time, the term $(\gamma \cdot \delta)$ represents learning rates  on a task,  and the term $(f + g \cdot \tau \cdot t^{a}) $ represents the growth in computation. So the overall term $((\gamma \cdot \delta) \cdot (\tau \cdot t^{a}) \cdot (f + g \cdot \tau \cdot t^{a}))$ represents the compound effect of learning rates $(\gamma \cdot \delta)$, price decline rates ($\tau \cdot t^{a}$), and growth in computation ($f + g \cdot \tau \cdot t^{a}$). Increases in any of these terms increase the rate of adoption. 

Equation \eqref{logit_updated4} provides an even simpler decomposition involving the compounding of the term ($\tau \cdot t^{a}$) that governs the rate at which AI becomes cheaper over time with the variable elasticity of substitution  $\sigma(\gamma, \delta, C(t))$.

Using the half-life to characterize a monotonic function often makes it easier to visualize. In the present case, a half-life of AI adoption occurs when $S_{\text{AI}}=0.5$. In \eqref{logit_updated4} this would lead to \eqref{logit_updated5}.

\begin{align}
0&=\ln(K) -  (\tau \cdot t^{a}) \cdot \sigma(\gamma, \delta, C(t)) \label{logit_updated5}
\end{align}

\noindent
A number of speculations center on the time-lines of AI adoption. Solving \eqref{logit_updated5} for $t$ gives $t_{\text{half}}$ in \eqref{logit_updated6}. 

\begin{align}
t_{\nicefrac12}&= \left [ \dfrac{\ln(K)}{\tau \cdot \sigma(\gamma, \delta, C(t))}  \right]^{1/a} \label{logit_updated6}
\end{align}

We present some values for $t_{\nicefrac12}$ in Table~\ref{thalf_tab}. This gives the half-life of adoption $t_{\nicefrac12}$ as a function of how bad the initial answer is $K$, the elasticity of substitution $\sigma$, and the speed of price declines $\tau$. These parameters could differ across tasks or products. For example, for a complicated query the initial answer could be quite poor (high $K$), there could be different AI products varying in price and the trajectory of price. If more involved AI products involve higher marginal costs, these could decline at a different rate than a small open source product. Finally, the elasticity of substitution will vary across AI tasks or products. For example, it may prove difficult by virtue of frictions to substitute AI for human products or tasks, especially if humans have been selected or matched to tasks that use their talents. We provide a range of estimates in Table~\ref{thalf_tab} across two orders of magnitude variation in $K$ and factor of four variation in $\sigma$, $\tau$. We see estimates of the half-life of adoption $t_{\nicefrac12}$ from 43.28 years to 1.55 years. The mean, median $t_{\nicefrac12}$ is 11.59, 8.52 years. Note, even if it takes a decade for AI for a particular task to obtain a 50\% market share, this is the equivalent to human labor for a particular  task or product falling by 6.9\% per year (continuous compounding), a  large change. 

Although some researchers predict AGI has already arrived in 2024, based on the ARC-AGI scores of the o3 model (Chollet, 2024), these AI adoption figures argue for a somewhat slower rate of adoption, much of which may come from a lower rate of substitution.\footnote{The ARC-AGI grand prize of one-million dollars will be awarded to a score of 85\% or better (given some other constraints such as computational cost). The Open AI o3 model scored above 85\%, but dramatically violated the computational constraints. Chollet (2024) considered this an impressive performance, but did not view it as AGI.} For example, despite the long-standing availability of effective image recognition technologies as well as the immense savings from employing these, the IRS will not be able to digitally process all paper documents until the 2026 filing season (IRS, 2023). We do not advocate for any of these particular scenarios, as stated earlier specific tasks or products in different settings may follow some of these scenarios, and we give the scenarios to allow readers to use their own priors.

%  The current high level in 2024 of 43\%. The organizers (arcprize.org) believe that current AI systems will only show slow increases in this benchmark over time. 

\setstretch{1}
\begin{table}[ptbh] 
\begin{center} 
\begin{tabular}[c]{l  r @{} l  r @{} l  r @{} l  r @{} l } 
\toprule 
\multicolumn{1}{l}{Case}& \multicolumn{2}{c}{$K$}& \multicolumn{2}{c}{$\sigma$}& \multicolumn{2}{c}{$\tau$}& \multicolumn{2}{c}{$t_{\nicefrac12}$}\\ 
\midrule 
1 & $500$&$.00$ & $0$&$.50$ & $0$&$.50$ & $24$&$.86$\\
2 & $500$&$.00$ & $0$&$.50$ & $1$&$.00$ & $12$&$.43$\\
3 & $500$&$.00$ & $0$&$.50$ & $2$&$.00$ & $6$&$.21$\\
4 & $500$&$.00$ & $1$&$.00$ & $0$&$.50$ & $12$&$.43$\\
5 & $500$&$.00$ & $1$&$.00$ & $1$&$.00$ & $6$&$.21$\\
6 & $500$&$.00$ & $1$&$.00$ & $2$&$.00$ & $3$&$.11$\\
7 & $500$&$.00$ & $2$&$.00$ & $0$&$.50$ & $6$&$.21$\\
8 & $500$&$.00$ & $2$&$.00$ & $1$&$.00$ & $3$&$.11$\\
9 & $500$&$.00$ & $2$&$.00$ & $2$&$.00$ & $1$&$.55$\\
10 & $5,000$&$.00$ & $0$&$.50$ & $0$&$.50$ & $34$&$.07$\\
11 & $5,000$&$.00$ & $0$&$.50$ & $1$&$.00$ & $17$&$.03$\\
12 & $5,000$&$.00$ & $0$&$.50$ & $2$&$.00$ & $8$&$.52$\\
13 & $5,000$&$.00$ & $1$&$.00$ & $0$&$.50$ & $17$&$.03$\\
14 & $5,000$&$.00$ & $1$&$.00$ & $1$&$.00$ & $8$&$.52$\\
15 & $5,000$&$.00$ & $1$&$.00$ & $2$&$.00$ & $4$&$.26$\\
16 & $5,000$&$.00$ & $2$&$.00$ & $0$&$.50$ & $8$&$.52$\\
17 & $5,000$&$.00$ & $2$&$.00$ & $1$&$.00$ & $4$&$.26$\\
18 & $5,000$&$.00$ & $2$&$.00$ & $2$&$.00$ & $2$&$.13$\\
19 & $50,000$&$.00$ & $0$&$.50$ & $0$&$.50$ & $43$&$.28$\\
20 & $50,000$&$.00$ & $0$&$.50$ & $1$&$.00$ & $21$&$.64$\\
21 & $50,000$&$.00$ & $0$&$.50$ & $2$&$.00$ & $10$&$.82$\\
22 & $50,000$&$.00$ & $1$&$.00$ & $0$&$.50$ & $21$&$.64$\\
23 & $50,000$&$.00$ & $1$&$.00$ & $1$&$.00$ & $10$&$.82$\\
24 & $50,000$&$.00$ & $1$&$.00$ & $2$&$.00$ & $5$&$.41$\\
25 & $50,000$&$.00$ & $2$&$.00$ & $0$&$.50$ & $10$&$.82$\\
26 & $50,000$&$.00$ & $2$&$.00$ & $1$&$.00$ & $5$&$.41$\\
27 & $50,000$&$.00$ & $2$&$.00$ & $2$&$.00$ & $2$&$.70$\\
\bottomrule 
\end{tabular} 
\caption{Half Life of AI Adoption as function of $K$, $\sigma$, and $\tau$} 
\label{thalf_tab} 
\end{center} 
\end{table}

\clearpage

\setstretch{1.25}

\section{Conclusion}

In an industry with an incumbent good with a stable price competing with an emergent good with a falling price, choices based on CES utility lead to the emergent good following either a logistic or a log-logistic S-curve of adoption.  The slope of the adoption S-curves depends on the interaction between the experience and elasticity of substitution parameters. The same relation holds when the prices between the emergent and incumbent goods remain the same, but the emergent good increases in quality which increases the elasticity of substitution in a VES setting.

These functional relations between price and adoption curves can provide building blocks for more complex models. The link between the price and adoption curves serves as a form of prior information that may allow for joint estimation of both series.  Because most of the interest in these series occurs in the early stages of adoption, such prior information may aid in forecasting future prices and adoption. More accurate forecasts crucially affect investment decisions, the design of innovation policies in areas such as renewable energy, artificial intelligence, and other emerging fields (Agrawal et al., 2019) as well as the trade-off between human and AI labor (Acemoglu et al., 2022). Deviations between the series may help identify non-price factors  impeding or accelerating adoption such as regulation.

%\section{Funding statment}

%The authors declare no funding.

\clearpage

\section*{References}

%\setstretch{1}

%\mybib Amir, Rabah, Philip Erickson, and Jim Jin (2007), ``On the Microeconomic Foundations of Linear Demand for Differentiated Products,'' \textit{Journal of Economic Theory}, Vol. 169, pp. 641-665.

%\mybib Arnold, Barry C. (2014), ``Univariate and Multivariate Pareto models,'' \textit{Journal of Statistical Distributions and Applications}, doi:10.1186/2195-5832-1-11.

%\mybib Arnold Barry C., Christopher A. Robertson and Hsiaw-Chan Yeh (1986), ``Some Properties of a Pareto-Type Distribution,'' \textit{Sankhy\=a: The Indian Journal of Statistics, Series A}, Vol. 48, No. 3  pp. 404--408.

\mybib Acemoglu, D., D. Autor, J. Hazell,  and P. Restrepo (2022), ``Artificial Intelligence and Jobs: Evidence from Online Vacancies,'' \textit{Journal of Labor Economics}, Vol. 40, S1, S293-S340.

%\mybib Adner, R., \& Kapoor, R. (2016). Innovation ecosystems and the pace of substitution: Re-examining technology S-curves. Strategic Management Journal, 37(4), 625--648.

\mybib Agrawal, A., and J. Gans, and A. Goldfarb (2019), ``Economic policy for artificial intelligence,'' \textit{Innovation policy and the economy}, Vol. 19, No. 1, pp.139--159.

\mybib Arrow, Kenneth J., H.B. Chenery, B.S. Minhas, and R.M. Solow (1961), ``Capital-labor Substitution and Economic Efficiency,'' \textit{Review of Economics and Statistics}, Vol. 43, No. 3, pp. 225--250.

\mybib Arrow, Kenneth J. (1962), ``The Economic Implications of Learning by Doing,'' \textit{The Review of Economic Studies}, Vol. 29, No. 3,  pp. 155--173.

\mybib Aslanyan, Anna (2024), ``AI Translation: How to Train the Horses of the Enlightenment,'' \textit{The Guardian}, March 15th.
%\mybib Balakrishnan, N. (1992), \textit{Handbook of The Logistic Distribution}, Marcel Dekker, New York.

%\mybib Bass, Frank (1969), ``A new product growth for model consumer durables,'' \textit{Management Science}, Vol. 15 No. 5, pp. 215--227. 

\mybib
Bass, F.M. (1969). A new product growth for model consumer durables. \textit{Management Science}, 15(5), 215--227.

\mybib Bass, Frank, Trichy V. Krishnan, and Dipak C. Jain (1994), ``Why the Bass Model Fits without Decision Variables,'' 
\textit{Marketing Science}, Vol. 13, No. 3, pp.  203--223.

\mybib Chollet, Fran\c{c}ois (2024), ``OpenAI o3 Breakthrough High Score on ARC-AGI-Pub,'' ARCprize.org.

\mybib Creutzig, Felix, J\'{e}r\^{o}me Hilaire, Gregory Nemet, Finn M\"{u}ller-Hansen, and Jan C. Minx (2023), ``Technological innovation enables low cost climate change mitigation,'' \textit{Energy Research \& Social Science}, 105,  103276.

\mybib Dixit, Avinash, and Joseph Stiglitz (1977), ``Monopolistic Competition and Optimum Product Diversity,'' \textit{American Economic Review}, Vol. 67, No. 3, pp.  297--308.

\mybib Erdil, Ege and Tamay Besiroglu (2024), ``Explosive Growth from AI Automation: A Review of the Arguments,'' ArXiv: 2039.11690v3.

\mybib Fisk, P.R. (1961), ``The Graduation of Income Distributions,'' \textit{Econometrica}, Vol. 29, pp. 171--185.

%\mybib G\o rtz, Erik (1977), ``An Identity between Price Elasticities and the Elasticity of Substitution of the Utility
%Function,'' \textit{The Scandinavian Journal of Economics}, Vol. 79, No. 4, pp. 497--499.

\mybib Griliches, Zvi (1957), ``Hybrid Corn: An Exploration in the Economics of Technological Change,'' \textit{Econometrica}, Vol. 25, No. 4, pp. 501--522.

\mybib Gr\"{u}bler, Arnulf, Neboj\v{s}a Naki\'{c}enovi\'{c}, and David G. Victor (1999), ``Dynamics of energy technologies and global change,'' \textit{Energy Policy}, Vol. 27, pp.  247--280.

%\mybib Johnson, N.L., S. Kotz, and N. Balakrishnan (1995), \textit{Continuous Univariate Distributions}, Vol. 2, Wiley, New
%York.

\mybib Haegel, Nancy M. and Sarah R. Kurtz (2023), ``Global Progress Toward Renewable Electricity: Tracking the Role of Solar (Version 3),'' \textit{IEEE Journal of Photovoltaics}, doi: 10.1109/JPHOTOV.2023.3309922.

%\mybib Houthakker, H.S. (1960), ``Additive Preferences,'' \textit{Econometrica}, Vol. 28, No. 2 , pp. 244--257.

\mybib Hoffmann, Jordan,  Sebastian Borgeaud, Arthur Mensch, Elena Buchatskaya, Trevor Cai, Eliza Rutherford, Diego de Las Casas, Lisa Anne Hendricks, Johannes Welbl, Aidan Clark, Tom Hennigan, Eric Noland, Katie Millican, George van den Driessche, Bogdan Damoc, Aurelia Guy, Simon Osindero, Karen Simonyan, Erich Elsen, Jack W. Rae, Oriol Vinyals, and Laurent Sifre (2022), ``Training Compute-Optimal Large Language Models,'' arXiv: 2203.15556.

\mybib IRS (2023), ``Taxpayers will have the option to go paperless for IRS correspondence by 2024 filing season, IRS to achieve paperless processing for all tax returns by filing season 2025.'' FS-2023-18.

%Kaplan paper has 1111 cites
%\bibitem{Kaplan2020}
\mybib Kaplan, Jared, Sam McCandlish, Tom Henighan, Tom B. Brown, Benjamin Chess, Rewon Child, Scott Gray, Alec Radford, Jeffrey Wu, and Dario Amodei (2020), ``Scaling Laws for Neural Language Models.'' arXiv:2001.08361.

\mybib Korinek, Anton and Donghyun Suh (2024), ``Scenarios for the Transistion to AGI,'' NBER, 32255.

%N. M. Haegel and S. R. Kurtz, "Global Progress Toward Renewable Electricity: Tracking the Role of Solar (Version 3)," in IEEE Journal of Photovoltaics, doi: 10.1109/JPHOTOV.2023.3309922.

%\mybib Houthakker, H.S. (1965), ``A Note on Self-Dual Preferences,'' \textit{Econometrica}, Vol. 33, No. 4, pp. 797--801.

\mybib  Lafond, François, Aimee Gotway Bailey, Jan David Bakker, Dylan Rebois, Rubina Zadourian, Patrick McSharry, and J. Doyne Farmer  (2018), 
``How well do experience curves predict technological progress? A method for making distributional forecasts,'' \textit{Technological Forecasting \& Social Change},
Vol. 128, pp. 104--117.
%ISSN 0040-1625,
%https://doi.org/10.1016/j.techfore.2017.11.001.
%(https://www.sciencedirect.com/science/article/pii/S0040162517303736)

%\mybib Layard, P.R.G., and A.A. Walters (1978), \textit{Microeconomic Theory}, McGraw-Hill, New York. 

\mybib Li, Xin, Ruidong Chang, Jian Zuo, and Yanquan Zhang (2023), ``How does residential solar PV system diffusion occur in Australia? -- A logistic growth curve modelling approach,'' \textit{Sustainable Energy Technologies and Assessments}, Vol. 56, 103060. 
%https://doi.org/10.1016/j.seta.2023.103060.

\mybib  Magee, A., S. Basnet, J.L. Funk, and  C.L. Benson (2016), ``Quantitative empirical trends in technical performance,''
\textit{Technological Forecasting \& Social Change}, Vol. 104, pp. 237--246.

\mybib  Maslej, Nestor, Loredana Fattorini, Raymond Perrault, Vanessa Parli, Anka Reuel, Erik Brynjolfsson, John Etchemendy,
Katrina Ligett, Terah Lyons, James Manyika, Juan Carlos Niebles, Yoav Shoham, Russell Wald, and Jack Clark (2024),
“The AI Index 2024 Annual Report,” AI Index Steering Committee, Institute for Human-Centered AI, Stanford
University, Stanford, CA.

%\mybib Muse, Abdisalam Hassan, Samuel M. Mwalili, and Oscar Ngesa1 (2021), ``On the Log-Logistic Distribution and Its Generalizations: A Survey,'' \textit{International Journal of Statistics and Probability}, Vol. 10, No. 3, doi:10.5539/ijsp.v10n3p93.

\mybib Nagy B., J.D. Farmer, Q.M. Bui, and J.E. Trancik (2013), ``Statistical basis for predicting technological progress,'' \textit{PLOS One},  Vol. 8, e52669.
%330 GS cites
%Jessika E Trancik at MIT with 5K+ cites

\mybib Nijsse, F.J.M.M., Jean-Francois Mercure, Nadia Ameli, Francesca Larosa, Sumit Kothari, Jamie Rickman, Pim Vercoulen and Hector Pollitt  (2023), ``The Momentum of the Solar Energy Transition,'' \textit{Nature  Communications} 14, 6542, https://doi.org/10.1038/s41467-023-41971-7.

\mybib Odenweller,  Adrian (2022), ``Climate mitigation under S-shaped energy technology diffusion: Leveraging
synergies of optimisation and simulation models,'' \textit{Technological Forecasting \& Social Change},  178,  121568. 

\mybib Ott, Simon, Adriano Barbosa-Silva, Kathrin Blagec, Jan Brauner, and
Matthias Samwald (2022), ``Mapping Global Dynamics of Benchmark
Creation and Saturation in Artiﬁcial Intelligence,'' \textit{Nature Communications}, doi:10.1038/s41467-022-34591-0.

\mybib Rogers, E.M. (2010). \textit{Diffusion of innovations}, Simon and Schuster.

\mybib Sahal, D. (1979), ``A theory of progress functions,'' \textit{AIIE Transactions}, Vol. 11, No. 1, pp. 23-29. doi: 10.1080/05695557908974396.

%Sahal, D. (1979) ‘A Theory of Progress Functions’, A I I E Transactions, 11(1), pp. 23–29. doi: 10.1080/05695557908974396.

\mybib Seal, Thomas (2024), ``DeepMind CEO Says Google Will Spend More Than \$100 Billion on AI,'' \textit{Bloomberg}. April 15.

\mybib Sevilla, Jaime et al. (2024), ``Can AI Scaling Continue Through 2030?,''  epochai.org,  https://epochai.org/blog/can-ai-scaling-continue-through-2030.

\mybib Silberberg, Eugene and Wing Suen (2017), \textit{The Structure of Economics, A Mathematical Analysis}, McGraw Hill India, Chennai.

\mybib Singh, Anuraag, Giorgio Triulzi, and Christopher L. Magee (2021), ``Technological improvement rate predictions for all technologies: Use of patent data and an extended domain description,'' \textit{Research Policy}, Vol. 50,  104294.

%\bibitem{soa2024}
\mybib SoA Policy Team (2024), ``SoA Survey Reveals a Third of Translators and Quarter of Illustrators Losing Work to AI,'' \textit{Society of Authors}.

\mybib Triulzi, Giorgio, Jeff Alstott, and Christopher L. Magee (2020), ``Estimating technology performance improvement rates by mining patent data,'' \textit{Technological Forecasting \& Social Change},  158, 120100.

%\mybib  U.S. DOE’s Vehicle Technologies Office (8/21/2023), ``Fact 1304 Dataset,'' U.S. Energy Information Administration, Electric Power Monthly, February 2023, Tables 6.5 and 6.6.

\mybib Way, Rupert, Matthew C. Ives,
Penny Mealy, and J. Doyne Farmer (2022), ``Empirically grounded technology forecasts
and the energy transition,'' \textit{Joule}, Vol. 6, pp. 2057--2082.

%\mybib Verhulst, Pierre-François (1838), ``Notice sur la loi que la population poursuit dans son accroissement,'' \textit{Correspondance Mathématique et Physique}, Vol. 10, pp.  113--121. 

%\mybib Verhulst, Pierre-François (1845), ``Recherches mathématiques sur la loi d'accroissement de la population,'' \textit{Nouveaux Mémoires de l'Académie Royale des Sciences et Belles-Lettres de Bruxelles}, Vol. 18, No. 8. 

%\mybib Verhulst, Pierre-François (1847), ``Deuxième mémoire sur la loi d'accroissement de la population,'' \textit{Mémoires de l'Académie Royale des Sciences, des Lettres et des Beaux-Arts de Belgique}, Vol. 20, pp. 1–32. 

%\bibitem{guardian2024}

%Hoffmann paper has 942 cites

%\end{thebibliography}

\end{document}